\documentclass[aps,showpacs,preprintnumbers,amsmath,amssymb]{revtex4}

\usepackage{graphicx}
\usepackage{dcolumn}
\usepackage{bm}

\begin{document}


\title{Pomeranchuk and other instabilities in the $t-t'$ Hubbard model
\\ at the Van Hove filling}

\author{V.~Hankevych$^{1,2}$}
\author{I.~Grote$^1$}
\author{F.~Wegner$^1$}%
\affiliation{\mbox{$^1$Institute for Theoretical Physics, University of Heidelberg, Philosophenweg 19, D-69120 Heidelberg, Germany}\\
$^2$Department of Physics, Ternopil State Technical University, 56 Rus'ka Str., UA-46001 Ternopil, Ukraine}


\date{May 10, 2002}

\begin{abstract}
We present a stability analysis of the two-dimensional $t-t'$ Hubbard model for
various values of the next-nearest-neighbor hopping $t'$, and electron concentrations close to the Van Hove filling by means of the flow equation method. 
For $t'\geq -t/3$ a $d_{x^2-y^2}$-wave
Pomeranchuk instability dominates (apart from antiferromagnetism at small $t'$). At $t'<-t/3$ the leading instabilities are a
$g$-wave Pomeranchuk instability and $p$-wave particle-hole instability in the triplet channel at temperatures $T<0.15t$, and an $s^*$-magnetic phase for $T>0.15t$; upon increasing the electron concentration the triplet analog of the flux phase occurs at low temperatures. Other weaker instabilities are found also.
\end{abstract}

\maketitle


In recent years the two-dimensional (2D) Hubbard model has been used~\cite{iz,sc} as the simplest model which maps the electron correlations in the copper-oxide planes of high-temperature superconductors since experimental data suggest that superconductivity in cuprates basically originates from the CuO$_2$ 
layers~\cite{tk}. Although in the high-temperature cuprate superconductors electron-electron interactions are strong some important features of these systems 
(in particular, antiferromagnetic and $d$-wave superconducting instabilities) are captured already by the 2D Hubbard model at weak to moderate Coulomb coupling.

Apart from the antiferromagnetism and $d_{x^2-y^2}$-wave superconductivity 
mentioned above (for review see~\cite{iz,sc,ha} and references therein), a few other instabilities related to symmetry-broken 
states~\cite{hm,hsg,foh,ikk,hs,hsr,gkw} in the 2D $t-t'$ Hubbard model with 
next-nearest-neighbor hopping $t'$ have been reported recently. Specially, much interest of researchers has been attracted by the case when the Fermi surface passes through the saddle points of the single particle dispersion (Van Hove filling). One of the instabilities found in such a case is 
a $d$-wave Pomeranchuk instability breaking the tetragonal symmetry of the Fermi surface, i.e. a spontaneous deformation of the Fermi surface reducing its symmetry to orthorhombic. It has been recently observed for small values of $t'$ from renormalization group calculations by Halboth and Metzner~\cite{hm}. They argued that the Pomeranchuk instability occurs more easily if the Fermi surface is close to the saddle points with a sizable $t'$ (reducing nesting which leads to antiferromagnetism). However, within their technique it is difficult to compare the strength of the Fermi surface deformation with other instabilities and to conclude which one dominates. The authors of Ref.~\cite{hsr} have investigated the interplay of $d$-density wave~\cite{na,clmn} and Fermi surface deformation tendencies with those towards $d$-wave pairing and antiferromagnetism by means of a similar temperature-flow renormalization group approach. They have found that the $d$-wave Pomeranchuk instability never dominates in the 2D $t-t'$ Hubbard model (even under the conditions mentioned above). 

On the other hand, Vollhardt et al.~\cite{vbh} showed that the $t'$-hopping term destroys the antiferromagnetic nesting instability at weak interactions in two and three dimensions, and supports the stabilization of metallic ferromagnetism in infinite dimensions away from half-filling. Therefore, one could expect also the stabilization of ferromagnetism by a sizable $t'$ in two dimensions.
Indeed, in the $t-t'$ Hubbard model on a 2D square lattice at week to moderate Coulomb coupling, a projection quantum Monte Carlo calculation with 
$20\times 20$ sites and the $T$-matrix technique~\cite{hsg}, a generalized random phase approximation including particle-particle scattering~\cite{foh} 
point towards a ferromagnetic ground state for large negative values of $t'/t$ in a density range around the Van Hove filling. Similar tendencies have been found by the authors of Ref.~\cite{ikk} within the renormalization group and parquet approaches. Honerkamp and Salmhofer recently studied~\cite{hs} the stability of this ferromagnetic region at finite temperatures by means of the temperature-flow renormalization group technique. They have found that ferromagnetic instability is the leading one at $t'<-0.33t$ and Van Hove filling with critical temperatures depending on the value of $t'$. When the electron concentration is increased slightly above the Van Hove filling, the ferromagnetic tendencies get cut off at low temperatures and a triplet $p$-wave superconducting phase dominates. However, they did not consider the Pomeranchuk instability (which could have the most favorable conditions to occur) and other ones apart from antiferromagnetism, $d$- and $p$-wave superconductivity and ferromagnetism. 

Therefore, the investigation of interplay and rivalry between the Pomeranchuk instability and ferromagnetism, and other phases in the 2D $t-t'$ Hubbard model at the Van Hove filling is a considerable task. We will consider the leading instabilities depending on the ratio $U/t$ (in all papers cited above it was fixed). The main goal of this paper is such a study. We report also a few new instabilities in a range of electron concentration around the Van Hove filling. 

We start from the Hamiltonian of the $t-t'$ Hubbard model 
\begin{eqnarray}
H=\sum_{{\bf k}\sigma}\varepsilon_{\bf k}c^\dagger_{{\bf k}\sigma}c_{{\bf k}\sigma}+
{U\over N}\sum_{{\bf k_1 k_1' \atop k_2 k_2'}}c^\dagger_{{\bf k_1}\uparrow}c_{{\bf k_1'}\uparrow}c^\dagger_{{\bf k_2}\downarrow}c_{{\bf k_2'}\downarrow}\delta_{{\bf k_1}+{\bf k_2},{\bf k_1'}+{\bf k_2'}},
\end{eqnarray}
where $\varepsilon_{\bf k}$ is the Bloch electron energy with the momentum ${\bf k}$, $c^\dagger_{{\bf k}\sigma} (c_{{\bf k}\sigma})$ is the creation (annihilation) operator for the electrons with spin projection $\sigma \in \{\uparrow,\downarrow\}$, $U$ is the local Coulomb repulsion of two electrons of opposite spins, $N$ is the number of lattice points, lattice spacing equals unity.

By means of the flow equation method~\cite{we} the Hamiltonian is transformed into one of molecular-field type. This Hamiltonian is calculated in second order
in the coupling $U$~\cite{gkw}.
Adopting the notations of Ref.~\cite{gkw}, the expression for the free energy has the form:
\begin{eqnarray}
\beta F={1\over N}\sum_{\bf k q} \beta U\left(1+{U\over t}
V_{{\bf k}, {\bf q}}\right)\Delta_{\bf k}^{\ast}\Delta_{\bf q}+
\sum_{\bf k}f_{\bf k}\Delta_{\bf k}^{\ast}\Delta_{\bf k}, \label{fe}
\end{eqnarray}
where the first term is the energy contribution and the second term is the entropy contribution, $\beta =1/(k_BT)$, $T$ is the temperature, $t$ is the hopping integral of electrons between nearest neighbors of the lattice, $V_{{\bf k}, {\bf q}}$ is effective second-order interaction (the factor $U^2/t$ has been extracted from it), 
$f_{\bf k}$ is an entropy coefficient, and $\Delta_{\bf k}$ are the order parameters. For example, 
$\Delta_{{\bf k}\sigma, {\bf -k}\sigma'}=\langle c_{{\bf k}\sigma}
c_{{\bf -k}\sigma'}\rangle = (\sigma_y)_{\sigma\sigma'}\Delta_{\bf k}^{s}+
\sum_{\alpha}(\sigma_y\sigma_{\alpha})_{\sigma\sigma'}
\Delta_{\bf k}^{t\alpha}$, where $\sigma_{\alpha}$ is a Pauli spin matrix 
($\alpha=x, y, z$), and $\Delta_{\bf k}^{s}\ (\Delta_{\bf k}^{t\alpha})$ is the singlet (triplet) amplitude. An expression similar to Eq.~(\ref{fe}) is 
obtained for particle-hole channels with the order parameters $\nu$ instead of $\Delta$. In this case, for example, we have $\nu_{{\bf k}\sigma,{\bf k}\sigma'}=\langle c^\dagger_{{\bf k}\sigma}c_{{\bf k}+{\bf Q}\sigma'}\rangle=
\nu_{\bf k}^{s}\delta_{\sigma,\sigma'}+\sum_{\alpha}\nu_{\bf k}^{t\alpha}
(\sigma_{\alpha})_{\sigma\sigma'}$ with ${\bf Q}=(\pi,\pi)$. All quantities of Eq.~(\ref{fe}) are defined in Ref.~\cite{gkw}. 

For a square lattice the single particle dispersion has the form:
\begin{eqnarray}
\varepsilon_{\bf k}=-2t(\cos k_x+\cos k_y)-4t'\cos k_x\cos k_y. \label{ek}
\end{eqnarray}
The spectrum~(\ref{ek})
contains Van Hove singularities in the density of states at the energy $\varepsilon_{VH}=4t'$ related to the saddle points of the 
Fermi surface at ${\bf k}=(0,\pm\pi)$ and $(\pm\pi,0)$. 
For $t'=0$ and half-filling the Fermi surface is nested $\varepsilon_{{\bf k}+{\bf Q}}=-\varepsilon_{\bf k}$, which leads to an antiferromagnetic instability for $U>0$. The nesting is removed for $t'/t\neq 0$.

We start from the symmetric state and investigate whether this state is stable against fluctuations of the order parameters $\Delta$ and $\nu$. As soon as a non-zero $\Delta$ or $\nu$ yields a lower free energy in comparison with the symmetric state with all vanishing $\Delta$ and $\nu$, then the symmetric state is unstable and the system will approach a symmetry broken state. This indicates a phase transition.

We perform numerical calculation on a square lattice with $24\times 24$ 
points in the Brillouin zone for the various representations under the point group $C_{4\nu}$. 
The representations of the even-parity states are one-dimensional. We denote them by $s_+=s_1,\
s_-=s_{xy(x^2-y^2)},\ d_+=d_{x^2-y^2},\ d_-=d_{xy}$. The odd-parity representation is two-dimensional, here simply denoted by $p$. 
Initially, such numerical calculations have been performed in 
Refs.~\cite{gkw,gr}, but they were sensitive to the lattice size at low temperatures. Here we use an improved scheme (for details see Ref.~\cite{hw}).

\begin{figure}
\includegraphics{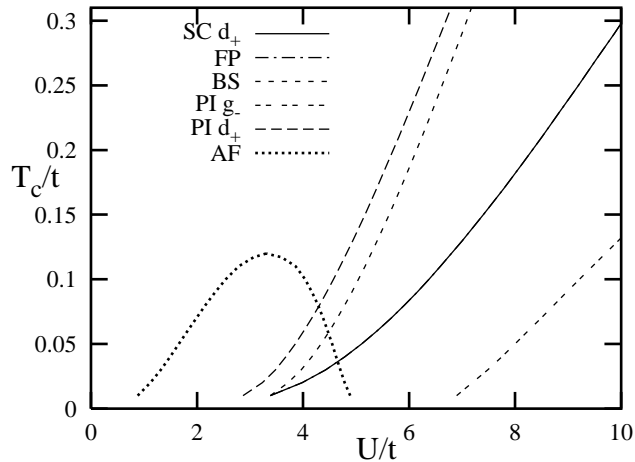}
\caption{\label{fig1} Temperature phase diagram of the 2D $t-t'$ Hubbard model for $n=1,\ t'=0$. Chemical potential $\mu=0$. SC stands for superconductivity, 
FP for flux phase, BS for band splitting, PI for Pomeranchuk instability, AF for antiferromagnetism.}
\end{figure}

\begin{figure}
\includegraphics{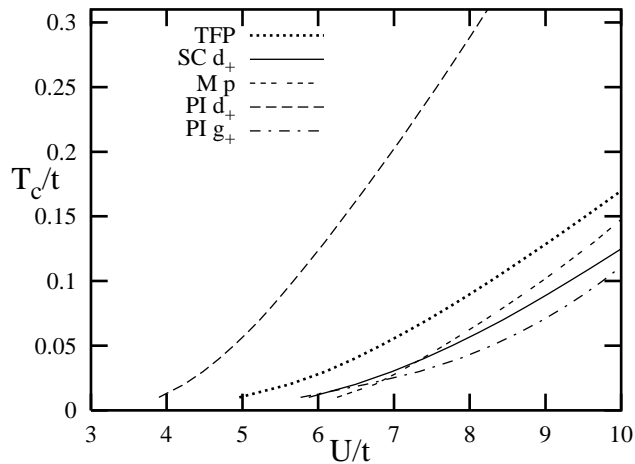}
\caption{\label{fig2} Temperature phase diagram of the 2D $t-t'$ Hubbard model for $t'=-t/3$, and $n=0.68$ (the Van Hove filling). Chemical potential varies between $\mu/t=-(1.317\div 1.339)$. TFP stands for triplet flux phase, M for magnetic particle-hole instability in triplet channel, other notations are the same as in Fig.~\ref{fig1}.}
\end{figure}

\begin{figure}
\includegraphics{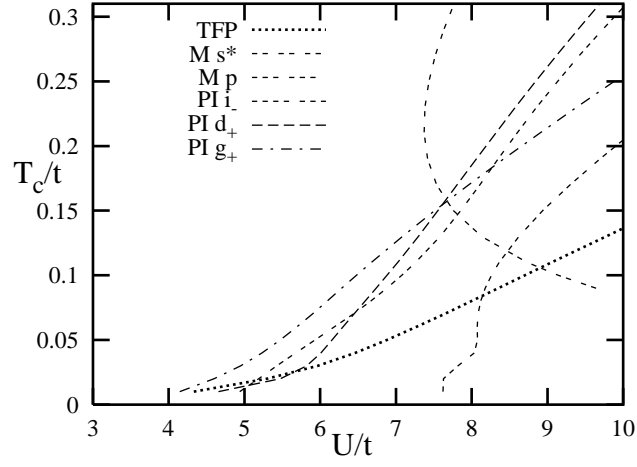}
\caption{\label{fig3} Temperature phase diagram of the 2D $t-t'$ Hubbard model for $t'=-5t/12$, and $n=0.55$ (the Van Hove filling). Chemical potential varies between $\mu/t=-(1.666\div 1.632)$. Notations are the same as in Fig.~\ref{fig2}.}
\end{figure}

We start from $t'=0$ and half-filling ($n=1$) (see Fig.~\ref{fig1}). 
As expected in this case the leading instability is the antiferromagnetic one which disappears at the temperature $T\approx 0.1t$ or doping $\delta\equiv n-1=0.06$. The next instability is a Pomeranchuk instability with $d_{x^2-y^2}$-wave symmetry in the singlet channel. The corresponding eigenvectors signals a deformation of the Fermi surface which breaks the point group symmetry of the square lattice. 
For negative $t'\geq -t/3$ the Pomeranchuk instability dominates at the Van Hove filling (see Fig.~\ref{fig2}). 
The $d_{x^2-y^2}$- wave Pomeranchuk instability
competes with other instabilities at $t'<-t/3$, and it is not the leading one (Fig.~\ref{fig3}).
In agreement with the ideas of Ref.~\cite{hm}
the instability is mainly driven by a strong attractive interaction between particles on opposite corners of the Fermi surface near the saddle points and a repulsive interaction between particles on neighboring corners. To favor such a behavior we need a sizable $t'$ reducing antiferromagnetic correlations. 

At half-filling and $t'=0$ the next instability is a particle-hole instability of singlet type with staggered $p$-wave symmetry. It yields~\cite{gkw} a splitting into two bands and may lead to an energy gap in the charge excitations spectrum. Another mechanism for a charge gap formation has been proposed~\cite{mj,klpt} recently in the 2D Hubbard model with $t'=0$ at weak coupling. The band splitting phase is developed in the region of electron concentration around half-filling, and is one of the strongest in that region. Then the superconducting $d_{x^2-y^2}$ instability follows which coincides with the $d_{x^2-y^2}$-wave staggered flux phase (the flux phase has been proposed by the authors of Ref.~\cite{am,ko} and discussed recently in Refs.~\cite{na,clmn,le}). Away from half-filling the degeneration disappears, and $d$-wave superconductivity dominates at low temperatures in certain regions of electron concentration around half-filling which depends on the value of $t'\neq 0$. Even large values of $|t'|$ do not destroy 
the dominant low-temperature behavior of $d_{x^2-y^2}$-wave superconductivity at doping~\cite{hw}. One phase may suppress another phase. To which extend two order parameters can coexist with each other is a question, which has to be investigated in the future. 

\begin{figure}
\includegraphics{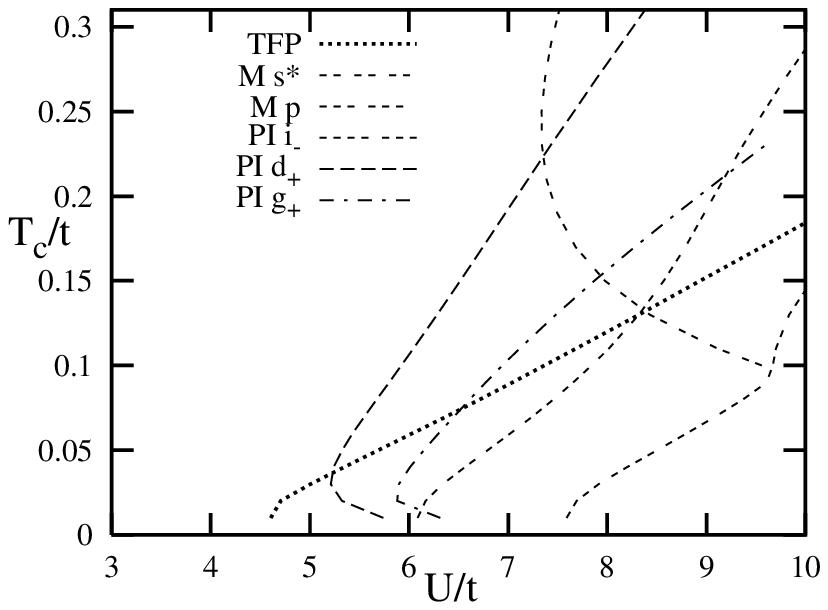}
\caption{\label{fig5} Temperature phase diagram of the 2D $t-t'$ Hubbard model for $t'=-5t/12$, and $n=0.60$ (slightly above the Van Hove filling). Chemical potential varies between $\mu/t=-(1.621\div 1.550)$. Notations are the same as in Fig.~\ref{fig2}.}
\end{figure}

For $t'=0$ the singlet and triplet $T_c$ 
of the particle-hole instabilities with staggered symmetry of $d_+$ wave character (that is the flux phase) are degenerate. If $t'\neq 0$ they are different, and the triplet one is higher. Moreover, the triplet analog of flux phase dominates at 
low temperatures and $t'=-5t/12$ when the electron concentration is slightly above the Van Hove filling (see Fig.~\ref{fig5}) in contrast to the results of Ref.~\cite{hs} which point out the occurrence of triplet superconductivity with $p$-wave symmetry in this region. The triplet flux phase is also one of the leading instabilities for $t'\geq -t/3$ and certain region of electron concentrations (see Fig.~\ref{fig2}). It has been considered by Nayak~\cite{na} as a density wave order parameter potentially relevant to the cuprates, but to our knowledge a triplet version of the flux phase has not yet been observed in numerical solutions of the 2D $t-t'$ Hubbard model. We shall discuss this state in more details elsewhere~\cite{hw}. 

\begin{figure}
\includegraphics{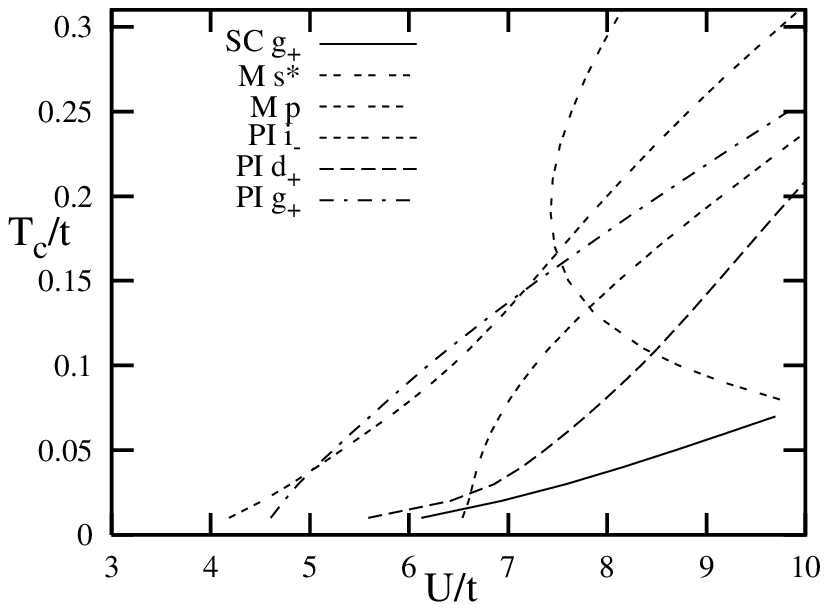}
\caption{\label{fig6} Temperature phase diagram of the 2D $t-t'$ Hubbard model for $t'=-5t/12$, and $n=0.50$ (slightly below the Van Hove filling). Chemical potential varies between $\mu/t=-(1.709\div 1.713)$. Notations are the same as in Fig.~\ref{fig2}.}
\end{figure}

At $t'=-5t/12$ a few other new instabilities appear to compete at the Van Hove filling and low temperatures (Fig.~\ref{fig3}) in disagreement with the conclusions of Ref.~\cite{hs} on the occurrence of ferromagnetism. The leading one is a Pomeranchuk instability in the $s_+$ channel with 
$g_+=g_{x^4+y^4-6x^2y^2}$ wave character (4 node lines in $\bf k$-space). 
This phase occurs more easily if the electron concentration is close to or slightly smaller than the Van Hove filling (Fig.~\ref{fig6}). It requests also sufficiently large absolute values of $t'$. When the electron concentration is decreased below the Van Hove density, a particle-hole instability of $p$-wave 
symmetry in triplet channel dominates at low temperatures (see Fig.~\ref{fig6}), which gives rise to a phase of magnetic currents. In the $d_-$ channel an $i$-wave (6 node lines in $\bf k$-space) Pomeranchuk instability appears when electron concentration $n$ is smaller or close to the Van Hove filling (Figs.~\ref{fig3}-\ref{fig6}). It is a leading one at small values of the electron 
concentration~\cite{hw}. We observe (Fig.~\ref{fig6}) in the $s_+$ channel a $g_+$ wave superconductivity below the Van Hove filling, but it requires strong coupling.

Another situation occurs at the temperature region $T>0.15t$. Here a particle-hole instability with 
$s^*$-wave character (its order parameter changes sign close to the Fermi-edge) in the triplet channel dominates at the density range around the Van Hove filling (see 
Figs.~\ref{fig3}-\ref{fig6}). It is likely that the order parameter contributions do not compensate exactly, so that a weak ferromagnetism appears.
When the electron concentration is increased above the Van Hove filling this instability does not become weaker, but the $d_{x^2-y^2}$ wave Pomeranchuk and the triplet flux phase instabilities are manifested stronger (they dominate at low temperatures). Then, this $s^*$-magnetic phase disappears at sufficiently large values of electron concentration in comparison with the Van Hove filling, or smaller $|t'|$. 

From Figs.~\ref{fig3}-\ref{fig6} one can see a reentrant behavior of the $s^*$-magnetic phase in some region of the values $U/t$: approaching $T_c$ we get $T_c^{l}$ from low temperatures and $T_c^{u}$ from high temperatures at the same value of coupling $U/t$ ($T_c^{l}\neq T_c^{u}$). This is a result of different behavior of $T_c(U/t)$ in two regimes. First regime occurs in the situation where the $s^*$-magnetic instability dominates and the transition from a 
paramagnetic state to the $s^*$-magnetic phase occurs directly without any intermediate phase, it corresponds to the temperatures $T>0.15t$ on Figs.~\ref{fig3},~\ref{fig6}. In this case the critical temperature increases with the increase of correlation strength $U/t$. 
Another regime occurs at the temperatures $T<0.15t$. 
In this situation the critical temperature exhibits an anomalous behavior, it decreases with increasing the coupling $U/t$. The $s^*$-magnetic phase is reduced. 
Since at lower temperatures only a smaller region in $\bf k$-space around the Fermi-edge contributes, the sign-change of the order parameter reduces the effective interaction. 

In conclusion, we have presented a stability analysis of the 2D $t-t'$ Hubbard model on a square lattice for various values of the next-nearest-neighbor hopping $t'$ and electron concentrations close to the Van Hove filling. A surprising large number of phases has been observed. Some of them have an order parameter with many nodes in $\bf k$-space. For $t'\geq -t/3$ the $d_{x^2-y^2}$-wave
Pomeranchuk instability dominates. At $t'<-t/3$ the leading instabilities are a
$g_+$ wave Pomeranchuk instability and $p$-wave particle-hole instability in triplet channel at temperatures $T<0.15t$, and $s^*$-magnetic phase for $T>0.15t$; upon increasing the electron concentration the triplet flux phase occurs at low temperatures. The $s^*$-magnetic phase is reduced strongly at low temperatures. We have found other weaker instabilities also.




\end{document}